\documentclass
[superscriptaddress,secnumarabic,nobibnotes,aps,prl,showkeys,noshowpacs,twocolumn,10pt]{revtex4-2}%
\usepackage{graphics}
\usepackage{graphicx}
\usepackage{subcaption}
\usepackage{epsf}
\usepackage{bm}
\usepackage{amsmath,amssymb,amsfonts,mathrsfs,amsthm}
\usepackage{latexsym}
\usepackage{enumerate}
\usepackage{comment}
\usepackage[dvipsnames,svgnames,x11names,table]{xcolor}
\usepackage[colorlinks = true,
            linkcolor = Cerulean,
            urlcolor  = magenta,
            citecolor = magenta,
            anchorcolor = NavyBlue]{hyperref}
\usepackage{epstopdf}%

\def\be{\begin{equation}}
\def\ee{\end{equation}}

\makeatother
\newcommand\RR{{\mathbb R}}

\begin{document}

\title{Quantum Mechanics from General Relativity and the Quantum Friedmann Equation}
\author{M. Matone}
\email{matone@pd.infn.it}
\affiliation{Dipartimento di Fisica e Astronomia ``G. Galilei'', Universit\`a di Padova, Via Marzolo 8 – 35131 Padova, Italy, Istituto Nazionale di Fisica Nucleare, sezione di Padova}
\author{N. Dimakis}
\email{nikolaos.dimakis@ufrontera.cl}
\affiliation{Departamento de Ciencias F\'{\i}sicas, Universidad de la Frontera, Casilla 54-D, 4811186 Temuco, Chile}

\begin{abstract}
We demonstrate that the recently introduced linear equation, reformulating the first Friedmann equation, is the first-order WKB expansion of a quantum cosmological equation.
This result shows a deeper underlying connection between General Relativity and Quantum Mechanics, pointing towards a unified framework.
Solutions of this equation are built in terms of a scale factor encapsulating quantum effects on a free-falling particle.
The quantum scale factor reshapes cosmic dynamics, resolving singularities at its vanishing points in several cases of interest.
As an explicit example, we consider the radiation-dominated era and show that the quantum equation is dual to the one in Seiberg-Witten formulation, recently applied to black holes, and incorporates resurgence phenomena and complex metrics, as developed by Kontsevich, Segal, and Witten. This links to the invariance of time parametrization under $\Gamma(2)$ transformations of the dual wave function.

\end{abstract}

\maketitle

\section{Introduction}

It is widely recognized that the domains of the physical world governed by gravity and those governed by quantum processes are challenging to reconcile \cite{Ashtekar:1974uu,Carlip}.
General Relativity (GR) is a highly non-linear theory, whereas
Quantum Mechanics (QM) is fundamentally based on linear equations, with the superposition principle playing a crucial role in describing quantum phenomena. Another often-stressed difference concerns the probabilistic interpretation of QM, in contrast with the deterministic structure of GR. However, there are compelling arguments suggesting a deep connection between GR and QM. An example arises in the loop expansion of the effective action in Quantum Field Theory, which assumes that the ratio of masses to
$\hbar$ is independent of $\hbar$.
In other words, there exists a relation of the form $m=\nu\hbar$, with $\nu$ being $\hbar$-independent.
For instance, in the case of a scalar field $\varphi$ with mass $m$ and scale $\mu$, we have
\begin{equation}
\Gamma[\varphi;\nu_m,\nu_\mu]=\sum_{k\geq0}\hbar^k
\Gamma_k[\varphi;\nu_m,\nu_\mu] \ ,
\label{QFT}\end{equation}
where $\nu_m:=\frac{m}{\hbar}$, $\nu_\mu: = \frac{\mu}{\hbar}$
are $\hbar$ independent (sect. 8.3 of \cite{QFT}).
As highlighted by Itzykson and Zuber, even for the Klein-Gordon equation, one considers (p. 288 of \cite{Itzykson:1980rh})
\begin{equation}
[\partial_x^2+(mc/\hbar)^2]\varphi=0 \ .
\label{KG}\end{equation}
Thus, dimensional analysis reveals that even at the classical level, a constant with the dimension of action is necessary, indicating that mass has a quantum origin. Since mass also serves as the gravitational charge, this suggests that QM and GR share a common foundation.

\noindent Here, we highlight key similarities between GR and QM emerging at the cosmological level. The reduction of GR to a system with finite degrees of freedom via cosmological configurations enables direct comparison with QM, leading to the framework of quantum cosmology \cite{QC1,QC2,QC3,QC4,QC5,QC6,QC7,QC8,QC9}. It is widely anticipated that a quantum framework for gravity may help resolve fundamental issues in modern cosmology \cite{PC1,PC2,PC3,PC4,PC5,PC6,PC7,PC8,PC9,sing1}.

\noindent In this setting, the Friedmann equations play a central role, describing the fundamental dynamics of cosmic evolution. While the second Friedmann equation is linear, the first is not. However, as shown in \cite{Matone:2021rfj}, it can be reformulated as a second-order linear
differential
equation that, for nonzero spatial scalar curvature  ${}^3R$, is unique. This naturally selects the conformal time, which plays a role analogous to the Hamilton's characteristic function.

We start from the key insight that redshift causes solutions of the linear equation to correspond to the first-order WKB approximation of a quantum cosmological equation. This correspondence uniquely yields the quantum Friedmann equation
\begin{equation}
\left(\frac{24}{c^2}\frac{d^2}{dt^2}+{}^3R(t)\right) \Psi(t)=0 \ ,
\label{bellissimaFrankZappa}\end{equation}
where ${}^3R(t)$ is explicitly expressed in terms of $t$, that is on shell.
A similar form holds in the case of vanishing spatial curvature.
Quantum effects are fully encoded in geometry, particularly in the quantum scale factor, which, in several cases of interest, resolves singularities at its zeros and enables continuous transitions between geometries.
We illustrate the formulation in the radiation-dominated case, where quantum solutions connect to the Seiberg-Witten (SW) approach: the Higgs vev equation matches the quantum Friedmann equation with $k \to -k$.
Here, complexified time exhibits a $\Gamma(2)$ invariance,
\begin{equation}
t(\tilde \Psi) = t(\Psi) \ ,
\end{equation}
suggesting analogies with particle physics phenomena like instantons and monopole condensation. These results tie into resurgence theory and complex metrics, as developed by Kontsevich, Segal, and Witten.

\section{Friedmann equations as Schwarzian equations}

Let us consider Einstein's equations,
\begin{equation}
  R_{\mu\nu}-\frac{1}{2} g_{\mu\nu}R +\Lambda g_{\mu\nu}=\frac{8\pi G}{c^4} T_{\mu\nu} \, ,
\end{equation}
where $\Lambda$ is the cosmological constant.
Assuming the Friedmann-Lemaître-Robertson-Walker metric
\begin{equation}
  ds^2 = -c^2dt^2 + a_k(t)^2 \left[ \frac{dr^2}{1-k r^2} +r^2 \left( d\theta^2 + \sin^2\theta d\varphi^2\right)  \right] \, ,
\end{equation}
and requiring that the matter sector shares its symmetries, i.e.\ a comoving,
homogeneous and isotropic perfect fluid of energy density $\rho$ and pressure $p$, with $u^\mu=(1,0,0,0)$, Einstein's
equations reduce to the two independent Friedmann equations \cite{Fried} 
\begin{subequations} \label{Freqs}
  \begin{align} \label{Freq1}
    F_1: \frac{\Lambda c^2}{3} -\frac{\kappa}{a_k^2} + \frac{8\pi G}{3} \rho - \frac{{\dot{a}_k}^2}{a_k^2} &=0 \ , \\ \label{Freq2}
    F_2: \frac{\Lambda c^2}{3} -\frac{4 \pi G}{c^2} \left( p + \frac{c^2 \rho}{3}  \right) - \frac{\ddot{a}_k}{a_k} & =0 \ ,
  \end{align}
\end{subequations}
where $\kappa=t_P^{-2}k=\ell_P^{-2}c^2k$, $k=-1,0,1$,
$t_P=\sqrt{\hbar G/c^5}$ the Planck time and $\ell_P$ is the Planck length.

Symmetries of the Friedmann equations have been extensively studied in \cite{Matone:2021rfj,Chimento,Faraoni,FrInfinite,Ben}.
Let
\begin{equation}
  \eta_k(t)=\int_{t}^{t_0} dt'a_k^{-1}(t') \, ,
\end{equation}
denote the conformal time, with $t_0$ today's value of proper time. If the lower integration bound is fixed at the beginning of the universe,
$\eta_k$ gives the conformal age of the universe, i.e.\ the particle horizon.
 A transformation from $t$ to $\eta_k$ brings the spacetime metric to the form
\begin{equation}
  ds^2 = a_k(\eta_k)^2 \left[ -c^2 d\eta_k^2 + \frac{dr^2}{1-k r^2} +r^2 \left( d\theta^2 + \sin^2\theta d\varphi^2\right)  \right] \, ,
\end{equation}
which makes apparent the conformal flatness of the spacetime for $k=0$. The $k\neq0$ case needs an additional spatial transformation \cite{conformalF}.

In \cite{Matone:2021rfj} it has been observed that, using the identity
\begin{equation}
\frac{\ddot a_k}{a_k}-\frac{1}{2}\left(\frac{\dot a_k}{a_k}\right)^2=-\{\eta_k,t\} \ ,
\label{addition1}\end{equation}
the linear combination
\begin{equation}
F_3=F_2-\frac{F_1}{2}=0 \ ,
\label{F3}\end{equation}
corresponds to
\begin{equation}
\frac{\kappa}{4}{\dot\eta_k}^2 - \frac{2}{3}\pi G\left(2\rho + \frac{3p}{c^2}\right) + \frac{\Lambda c^2}{12}
+\frac{1}{2}\{\eta_k,t\}=0 \ ,
\label{421}\end{equation}
where $\{\eta_k,t\}:=\dddot\eta_k/\dot\eta_k-3(\ddot\eta_k/\dot\eta_k)^2/2$ is the Schwarzian derivative of $\eta_k$.
In the case $k=0$ we then have
\begin{equation}
\frac{1}{2}\{\eta_0,t\}=\frac{2}{3}\pi G\left(2\rho + \frac{3p}{c^2}\right) - \frac{\Lambda c^2}{12} \ ,
\label{schwarzian2}\end{equation}
and, for $k=\pm 1$,
\begin{equation}
\frac{1}{2}\{e^{i\sqrt{\kappa}\eta_k},t\}=\frac{2}{3}\pi G\left(2\rho + \frac{3p}{c^2}\right) - \frac{\Lambda c^2}{12} \ ,
\label{schwarzian3}\end{equation}
where we used the basic identity
\begin{equation}
\{e^{i\sqrt{\kappa}\eta_k},t\}=\{\eta_k,t\}+\frac{\kappa}{2}\dot\eta_k^2 \ .
\end{equation}
It follows that the two Friedmann equations can be replaced by \eqref{schwarzian2}, for $k=0$,
and by \eqref{schwarzian3} for $k=\pm1$, together with any other nontrivial linear combination $A F_1 + B F_2 = 0$, with constants $A$ and $B$ such that $A/B \neq -1/2$.

\noindent
As shown in \cite{Matone:2021rfj}, even the second Friedmann equation is equivalent to a Schwarzian equation. Namely,
\begin{equation}
\{t_1,t\}=\frac{8}{3}\pi G\left( \rho(t) + \frac{3p(t)}{c^2} \right) - \frac{2\Lambda c^2}{3} \ ,
\label{SchwarzF2}\end{equation}
 where
\begin{equation}
t_\beta(t)=\int^t dt'\,a_k^{-2\beta}(t') \ ,
\end{equation}
are the $\beta$-times.
Explicitly expressing $\rho(t) + \frac{3p(t)}{c^2}$ as a function of $t$, Eq.~\eqref{SchwarzF2} becomes a Schwarzian equation.

\noindent
We then have that the Friedmann equations are equivalent to the Schwarzian equations.
In particular, for $k=0$, the Friedmann equations are equivalent to \eqref{schwarzian2} and \eqref{SchwarzF2}, whereas
in the case $k=\pm 1$, they are equivalent to \eqref{schwarzian3} and \eqref{SchwarzF2}.

\noindent Later we will show that for $k=\pm1$, the equations \eqref{schwarzian3} and \eqref{SchwarzF2} are the unique Schwarzian equations equivalent to the two Friedmann equations.

\noindent
In the case $k=0$, there are, besides \eqref{SchwarzF2}, infinitely many equivalent Schwarzian equations \cite{Matone:2021rfj}.

\noindent
A well known property of the Schwarzian equations,
used above and, for example, in the case of the Quantum Stationary Hamilton-Jacobi Equation (QSHJE) \cite{Faraggi:1998pd}, is that the solutions of the highly
non-linear ODE
\begin{equation}
\{f(t),t\}=g(t) \ ,
\label{1S}\end{equation}
are given by
\begin{equation}
f(t)=\frac{A \phi_1(t)+B\phi_2(t)}{C\phi_1(t)+D\phi_2(t)} \ ,
\end{equation}
where the constants $A,B,C$ and $D$ must satisfy the condition $AD-BC\neq 0$, and $\phi_1$ and $\phi_2$ are
two arbitrary linearly independent solutions of the second-order linear differential equation
\begin{equation}
\left(\frac{d^2}{dt^2}+\frac{g(t)}{2}\right)\phi(t)=0 \ .
\label{2S}\end{equation}
Note that this also shows the invariance of the Schwarzian derivative under the M\"obius transformations
of $f(t)$.
We then have that solving the Schwarzian equations is equivalent to solve the associated second-order linear differential
equation.

\noindent
In the case of the stationary Schr\"odinger equation
\begin{equation}
\left(-\frac{\hbar^2}{2m}\frac{d^2}{dx^2}+V(x) - E\right)\phi(x)=0 \ ,
\label{Schroedinger}\end{equation}
setting $\phi=Re^{iS/\hbar}$, one gets the QSHJE
\begin{equation}
\frac{1}{2m}\left(\frac{d S}{dx}\right)^2+V(x)-E+Q(x)=0 \ ,
\end{equation}
together with the continuity equation
\begin{equation}
\frac{d}{dx}\left(R^2 S'\right)=0 \ ,
\label{CE}\end{equation}
where
\begin{equation}
Q=-\frac{\hbar^2}{2m}\frac{R''}{R} \ ,
\end{equation}
is the quantum potential.
Then, observing that by \eqref{CE}
\begin{equation}
R(x)= \frac{A}{\sqrt{S'(x)}} \ ,
\end{equation}
with $A$ an arbitrary constant, it follows that the quantum potential can be equivalently expressed in
the form
\begin{equation}
Q(x)= \frac{\hbar^2}{4m}\{S,x\} \ ,
\label{QuantumPotential}\end{equation}
that plays the role of
particle's self-energy.
We then have that the QSHJE can be expressed as a unique differential equation
\begin{equation}
\frac{1}{2m}\left(\frac{d S}{dx}\right)^2+V(x)-E+\frac{\hbar^2}{4m}\{S,x\}=0 \ .
\label{intqhje}\end{equation}
Also note that, due to the reality of the Schr\"odinger equation, it follows that
if $\Psi$ is a solution then $\Psi^D=\bar\Psi$ is still a solution, which is
linearly independent of $\Psi$, unless it is proportional to a real function.
Also note that by Wronskian arguments one may easily show that given a solution $\Psi$,
any other linearly independent solution has the form $\Phi(x)=\Psi(x)+ A\int^x dx'\Psi^{-2}(x')$, with $A$
a non-vanishing constant.
It follows that any solution of the Schr\"odinger equation is a linear combination of the two linearly independent solutions
\begin{equation}
\Psi=\frac{1}{\sqrt{S'}}e^{-iS/\hbar} \ , \qquad \Psi^D=\frac{1}{\sqrt{S'}}e^{iS/\hbar} \ .
\end{equation}
Let us consider the Schwarzian equation associated with the Schr\"odinger equation \eqref{Schroedinger}
\begin{equation}
\left\{ e^{2iS/\hbar}, x \right\} = -\frac{4m}{\hbar^2}(V - E) \ ,
\label{ODE}\end{equation}
and observe that, while this equation is invariant under the M\"obius transformation
\begin{equation}
e^{2iS/\hbar} \rightarrow e^{2i\tilde S/\hbar} = \frac{A e^{2iS/\hbar} + B}{C e^{2iS/\hbar} + D} \ ,
\end{equation}
this invariance does not extend to the individual terms $\{S, x\}$ and $(S')^2$. In fact, although the combination
$\{e^{2iS/\hbar},x\}=\{S,x\} + \frac{2}{\hbar^2} (S')^2$ remains invariant, both $\{S,x\}$ and $(S')^2$ transform nontrivially under such a map.
Naturally, these transformations correspond to different solutions of the Schr\"odinger equation.

\noindent
Let us now show how  the solutions of the QSHJE, equivalent to the non-linear ODE \eqref{ODE}, are obtained.
First, from the above remarks,
it follows that it is always possible to choose two real linearly independent solutions
of the Schr\"odinger equation, that we denote by $\phi(x)$ and $\phi^D(x)$.
It follows that given a real solution $\phi$ of the Schr\"odinger equation, there is always
an associated real $S$, namely
\begin{equation}
\phi=\frac{1}{\sqrt{S'}}\left(A e^{-\frac{i}{\hbar}S}+B e^{\frac{i}{\hbar}S}\right) \ ,
\end{equation}
where $B=\bar A$. Note that if $S'<0$, then $i\phi$ is real. The  expression of $\phi^D$
is the one of $\phi$ with $B$ replaced by $-B$.
Then the phase can be expressed as (for more details see sect. 14.4 of \cite{Faraggi:1998pd})
\begin{equation}
e^{\frac{2i}{\hbar}S}=e^{i\alpha}\frac{w+i\bar\ell}{w-i\ell} \ ,
\label{KdT3}\end{equation}
where  $w=\phi^D/\phi\in\mathbb{R}$, $\alpha$ is a real integration constant and $\ell=\ell_1+i\ell_2$, $\ell_1, \ell_2\in\mathbb{R}$, is a constant with the condition $\ell_1\neq 0$. Note that the inverse of \eqref{KdT3} reads
\begin{equation}
w=i \frac{\ell e^{\frac{2i}{\hbar}S}+\bar\ell e^{i\alpha}}{e^{\frac{2i}{\hbar}S}-e^{i\alpha}} \ .
\end{equation}
By \eqref{KdT3} the conjugate momentum then is
\begin{equation}
S'=\frac{\hbar W (\ell+\bar\ell)}{2\left|\phi^D-i\ell\phi\right|^2} \ ,
\label{momentino}\end{equation}
where $W=\phi{\phi^D}'-\phi'{\phi^D}\in\RR\backslash\{0\}$ is the Wronskian.

\section{Linear form of the Friedmann equations}

Let us now go back to Eq.~\eqref{421}. A key point of our construction is that, as observed in \cite{Matone:2021rfj}, the term $\frac{1}{2}\{\eta_k,t\}$ has the same form as the quantum potential, which is a characteristic property of the QSHJE.
This suggests that the Friedmann equations are related to QM.
Next, note that (as illustrated above), a direct inspection shows Eq.~\eqref{421} is equivalent to a linear differential equation. In particular, besides the second Friedmann equation (which is already a linear ODE for $a_k$), the first Friedmann equation can also be replaced by an equivalent linear form \cite{Matone:2021rfj}. Thus, we have
\begin{equation}
  \left[\frac{d^2}{dt^2} + \frac{2}{3}\pi G\left(2\rho(t)+ \frac{3p(t)}{c^2} \right)-\frac{\Lambda c^2}{12}\right] \psi_k(t)=0 \ ,
\label{lineqbhalf}\end{equation}
\begin{equation}
\left[\frac{d^2}{dt^2} +\frac{4}{3}\pi G\left( \rho(t) + \frac{3p(t)}{c^2} \right) - \frac{\Lambda c^2}{3}\right]a_k(t)=0 \ ,
\label{F2bis}\end{equation}
where, in terms of $a_k$, the solutions of Eq.~\eqref{lineqbhalf} are the linearly independent functions
\begin{equation}
\psi_k= \sqrt{a_k}e^{-\frac{i}{2}\sqrt{\kappa}\eta_k} \ , \,
\psi^D_k =\sqrt{a_k}e^{\frac{i}{2} \sqrt{\kappa}\eta_k} \ , \, k=\pm 1 \ ,
\label{lesoluzioni}\end{equation}
and, for $k=0$,
\begin{equation}
\psi_0 =\sqrt a_0 \ , \qquad \psi_0^D=\sqrt{a_0}\eta_0/t_P \ .
\label{lesoluzioni00}\end{equation}
Let us recall that performing the second derivative with respect to $t$ in \eqref{lineqbhalf}
we get \eqref{421}. Moreover, by \eqref{F3} we have
$F_1=2(F_2+F_3)=0$,
so that the above linear system is completely equivalent to the Friedmann equations.
This remains true once $\rho(t)$ and $p(t)$ are evaluated on shell.

\noindent
As already noted, the above linear system is equivalent to the one
where \eqref{F2bis} is replaced by any nontrivial linear combination $AF_1+BF_2=0$, with
$A\neq 0$.
For $k=\pm1$, any choice with $A/B\neq -1/2$ yields an equation that is not a second-order
linear differential equation, as we shall demonstrate shortly.

\noindent
Once $2\rho(t)+3p(t)/c^2$ in Eq.~\eqref{lineqbhalf}
is given by its explicit expression in terms of $t$,
one obtains time-dependent solutions which, when compared with \eqref{lesoluzioni} (for $k=\pm1$)
and \eqref{lesoluzioni00} (for $k=0$), yield the expression for $a_k(t)$, which is further constrained by Eq.~\eqref{F2bis}.
We will explicitly illustrate this procedure below.

\noindent
We emphasize that, by construction, Eq.~\eqref{lineqbhalf}
together with \eqref{F2bis}, encapsulates the full content
of the two Friedmann equations.
The key physical novelty lies in the fact that Eq.~\eqref{lineqbhalf}
unveils a deep connection with quantum mechanics
that remains hidden in the standard formulation.

\noindent
As observed in \cite{Matone:2021rfj}, we have the following result
that reveals the deep structural rigidity of the system.

\bigskip

\noindent
{\bf Proposition.}
{\it For $k = \pm 1$, other than \eqref{F2bis}, Eq.~\eqref{lineqbhalf} is the only second-order linear differential equation derivable from the Friedmann equations. Furthermore, for $k=\pm 1$, the equations \eqref{schwarzian3} and \eqref{SchwarzF2} are the only possible Schwarzian equations which are equivalent to the two Friedmann equations.}

\bigskip

\noindent
The key point in proving this is that in the linear combination $AF_1+BF_2=0$, unless $A=0$,
the curvature term $k/a_k^2$ always appears. On the other hand, such a term can be absorbed
into the Schwarzian derivative only in the
case $A/B=-1/2$ \cite{Matone:2021rfj}. This implies that only in this case we have a Schwarzian equation, and therefore an associated second-order linear differential equation,
that is Eq.~\eqref{lineqbhalf}. The rest of the proof follows from the correspondence between Schwarzian equations and
second-order linear differential equations, reported in Eqs.~\eqref{1S}-\eqref{2S}.

\noindent
In the case $k=0$,
there are infinitely many second-order linear differential equations which are equivalent to the Friedmann equations. This also implies that for $k=0$ there
are infinitely many hidden symmetries \cite{Matone:2021rfj}.

\noindent
Using the above findings, we now express $a_k$ in terms of the solutions of \eqref{lineqbhalf}.
Notice that
\begin{equation}
\frac{\psi^D_k}{\psi_k} = e^{i\sqrt{\kappa}\eta_{k}} \ ,
\label{fracpsiq}\end{equation}
which is a phase factor for $k=1$ and is real for $k=-1$.
It is clear that $a_{1}$ is the analogue of the inverse of $S'$, with $\hbar$
replaced by $2t_P$. Thus
\begin{equation}
a_{1} = \frac{|\phi_1^D-i \ell \phi_1|^2}{W t_P  (\ell+\bar{\ell})} \ ,
\label{a1}\end{equation}
where $\phi_1$ and $\phi_1^D$ are linearly independent real solutions of \eqref{lineqbhalf} and
$W=\phi_1\dot\phi_1^D-\dot\phi_1\phi_1^D\in\RR\backslash\{0\}$ is the Wronskian.
Note that, as expected, due to the properties of the Wronskian, $a_1$ is invariant under a rescaling of $\phi_1$ and $\phi_1^D$.

\noindent
Let us now consider the case $k=-1$. Since $\psi_{-1}$ and $\psi_{-1}^D$ are real, we turn to
\eqref{fracpsiq} and express $\psi_{-1}$ and $\psi_{-1}^D$ as a general linear combination of real $\phi_{-1}$ and $\phi_{-1}^D$
\begin{equation}
  e^{-\eta_{-1}/t_P} =\frac{\psi_{-1}^D}{\psi_{-1}} = \frac{\gamma \phi_{-1}^D + \delta \phi_{-1}}{\alpha \phi_{-1}^D + \zeta \phi_{-1}} \ ,
\end{equation}
leading to
\begin{equation} \label{aqm1}
a_{-1} = \frac{\left(\alpha \phi_{-1}^D + \zeta \phi_{-1}\right)\left( \gamma \phi_{-1}^D + \delta \phi_{-1} \right)}{W t_P   \left(\zeta\gamma-\alpha \delta\right)} \ ,
\end{equation}
where $\alpha,\zeta,\gamma,\delta \in \mathbb{R}$ with $\alpha \delta-\zeta\gamma\neq0$.

\noindent
In the case $k=0$, by \eqref{lesoluzioni00} we have
\begin{equation}
\eta_0= \frac{\psi_0^D}{t_P\psi_0}=\frac{A\phi_0^D+B\phi_0}{C\phi_0^D+D\phi_0} \ ,
\end{equation}
where $\phi_0$ and $\phi_0^D$ are linearly independent real solutions of \eqref{lineqbhalf}.
Computing the inverse of the derivative we get
\begin{equation} \label{azero}
a_0=t_P \frac{(C\phi_0^D+D\phi_0)^2}{W(AD-BC)} \ ,
\end{equation}
where $W=\phi_0 \dot\phi_0^D - \phi_0^D\dot\phi_0$ is the Wronskian.

\noindent
Now note that we have expressed $a_k$ in terms of real solutions of \eqref{lineqbhalf}. On the other hand, $a_k$ must also satisfy Eq.~\eqref{F2bis}, which leads to the constraints
\begin{equation}
O(t)|\phi_1^D-i \ell \phi_1|^2=0 \ ,
\label{F2bis1}\end{equation}
\begin{equation}
O(t){\left(\alpha \phi_{-1}^D + \zeta
\phi_{-1}\right)\left( \gamma \phi_{-1}^D + \delta \phi_{-1} \right)}=0 \ ,
\label{F2bis-1}\end{equation}
\begin{equation}
O(t)(C\phi_0^D+D\phi_0)^2=0 \ ,
\label{F2bis0}\end{equation}
where
\begin{equation}
O(t)=\frac{d^2}{dt^2} +\frac{4}{3}\pi G\left( \rho + \frac{3p}{c^2} \right) - \frac{\Lambda c^2}{3} \ .
\label{Ot}\end{equation}
The resulting scale factor obtained through this process is the one satisfying the original Friedmann system since both Eqs.~\eqref{lineqbhalf} and \eqref{F2bis} have been utilized. Also note that the highly non linear coupling between $\rho(t)$, $p(t)$ and $a(t)$ explicitly appears
in Eqs.~\eqref{a1} - \eqref{Ot}.

\noindent
As a simple example, we consider a perfect fluid
with a linear barotropic equation of state, i.e. $p=w c^2 \rho$, with $k=0$. In this
case the energy density decays as the inverse square of the cosmic time when $w \neq -1$.
It is straightforward to verify that
\begin{equation}
  \rho(t) = \frac{1}{6 \pi  G} \frac{1}{[(w+1)(t-t_0)]^2} \ ,
\end{equation}
where $t_0$ represents some initial time. Equations \eqref{lineqbhalf} and \eqref{F2bis} then read
\begin{align} \label{example1}
  \left[\frac{d^2}{dt^2} + \frac{3 w+2}{9 (w+1)^2 (t-t_0)^2} \right] \psi_0(t)=0\ , \\ \label{example2}
  \left[\frac{d^2}{dt^2} + \frac{2 (3 w+1)}{9 (w+1)^2 (t-t_0)^2}\right] a_0(t)=0 \ .
\end{align}
Solving the first we obtain
\begin{equation}
   \psi = A (t-t_0)^{\frac{1}{3 w+3}}+ B (t-t_0)^{\frac{3 w+2}{3 w+3}} \ ,
\end{equation}
when $w\neq -1/3$ and
\begin{equation}
   \psi = A \sqrt{t-t_0}+ B \sqrt{t-t_0} \log (t-t_0) \ ,
\end{equation}
if $w=-1/3$. In both cases, if we use \eqref{azero}, which implies $a_0 \sim \psi^2$ in the second equation, Eq.~\eqref{example2}, we deduce $B=0$. This results to the well known expression
\begin{equation}
  a_0 \sim (t-t_0)^{\frac{2}{3 (w+1)}} \ ,
\end{equation}
for a spatially flat universe in the case of a linear equation of state.

\section{Redshift and WKB approximation}

A possible interpretation of Eq.~\eqref{lineqbhalf} would be to consider it as the analog of the Schr\"odinger equation. On the other hand, the corresponding classical equation of motion would then be given by \eqref{421} without the term $\{\eta_k,t\}/2$, resulting in an equation that lacks an obvious classical interpretation.
Instead, this equation indicates that the solutions
\eqref{lesoluzioni} are the first-order WKB expansion of solutions of a quantum equation.
The key point emerges from the redshift relation
$\lambda(t)a(t_0)=\lambda(t_0)a(t)$, which extends
to the momentum of free-falling massive particles, so that
$P_m\sim a^{-1}$ \cite{Weinberg}, where
\begin{equation}
P_m=m\sqrt{g_{ij}\frac{dx^i}{d\tau}\frac{dx^j}{d\tau}} \ ,
\end{equation}
$d\tau^2= - c^2 dt^2 + g_{ij}dx^i dx^j$.
This relation also holds when $m\to 0$, $d\tau \to 0$, as is the case for photons, so that
the Planck constant emerges as a consistency condition, given by
$P_\gamma=h/\lambda_\gamma$.
Another suggestive relation between GR
and QM is that, as shown in \cite{Misner:1973prb}, the redshift relation applies to the de Broglie wavelength $P_m=h/\lambda_m$ as well.
Therefore, for a free-falling particle we have
\begin{equation}
a_k(t)P_k(t)=a_k(t_0)P_k(t_0) \ .
\label{aP}\end{equation}
A deeper connection between GR and QM follows from Eqs.~\eqref{lineqbhalf}, \eqref{lesoluzioni} and \eqref{aP}.
Namely, choosing the normalization
\begin{equation}
a_k(t_0)P_k(t_0)=\frac{1}{2}\sqrt{\hbar c^3/G} \ ,
\end{equation}
we have
\begin{equation}
\psi_k=\frac{1}{\sqrt {P_k}}e^{-i\frac{c\sqrt k}{\hbar}\int_t^{t_0} P_k} \ , \quad \psi_k^D=\frac{1}{\sqrt {P_k}}
e^{i\frac{c\sqrt k}{\hbar}\int_t^{t_0} P_k} \ ,
\label{lesoluzioniQM+1}\end{equation}
$k=\pm 1$. In QM, setting $\psi=\exp (i\sigma/\hbar)$, the WKB expansion $\sigma=\sum_{j\geq 0}(\hbar/i)^j\sigma_j$
at first-order in $\hbar$ reads
\begin{equation}
\sigma=S+\frac{i}{2}\hbar \log S' \ ,
\end{equation}
with $S=\pm\int^x dx' \sqrt{2m(E-V)}$ the Hamilton's characteristic function. Therefore, at first-order
\begin{equation}
\psi_{WKB}=P^{-1/2}\exp(i\int P/\hbar) \ ,
\label{psiWKB}\end{equation}
with $P=\pm\sqrt{2m(E-V)}=S'$ the classical momentum.

\section{Quantum cosmological equation}

The above analysis shows that the solutions of \eqref{lineqbhalf} have the same structure as the Schr\"odinger
equation, with the spatial coordinate replaced by $t$. This suggests interpreting $|\psi_k|^2 \sim 1/P_k$
as the quasi-classical probability density of detecting a free-falling particle within the time interval
$[t,t+dt]$. In this respect, note that the absence of spatial variables removes the issue of the non-positive
definiteness of the probability density that arises, for example, in the Klein--Gordon equation due to the
Minkowskian signature.

\noindent
A key point is that the formulation is closely related to the QSHJE as developed in
\cite{Faraggi:1997bd,Faraggi:1998pd}. This equation follows from an equivalence postulate requiring
the Hamilton's characteristic function to be invariant under point transformations. Such a requirement
imposes a cocycle condition, which in turn defines the Schwarzian derivative and implies the addition
of the quantum potential to the classical HJ equation. The main features include tunnelling, with
positive definiteness of the conjugate momentum across the barrier, energy quantisation without
any probabilistic interpretation of the wave function, and the absence of trajectories in compact spaces
\cite{Faraggi:2012fv}.

\noindent
We also note that \eqref{lineqbhalf} suggests the correspondence
\begin{equation}
P \sim \frac{d}{dt}\ ,
\label{bellatoponza}
\end{equation}
where determining the exact correspondence is linked to the scale invariance of the Friedmann
equations under $a_k \to \lambda a_k$, which persists even for $\kappa \neq 0$ through a common
rescaling of $a_k$ and $\sqrt{\kappa}$.

\noindent
We now derive the exact equation whose solutions reproduce \eqref{lesoluzioni} and \eqref{lesoluzioni00} to first-order in $t_P$.
As noted above, such an equation can be directly derived by observing that, by $a_k \sim 1/P_k$, the analogous of
\begin{equation}
\left(-\hbar^2 \frac{d^2}{d x^2}+P^2\right)\psi=0 \ ,
\end{equation}
is just
\begin{equation}
\left(\frac{d^2}{dt^2}+\frac{\alpha}{a_k^2}\right) \Psi = 0 \ ,
\label{generale}\end{equation}
with $\alpha$ a constant, that we fix in the case $k=\pm 1$.
We first refer to the well-known QM case.
Define $X$ in such a way that the solution of the Schr\"odinger
equation at the first order in $\hbar$, $\Psi_{WKB}$, satisfies
\begin{equation}
\left(-\frac{\hbar^2}{2m}\frac{d^2}{dx^2}+V-E+X\right)\psi_{WKB}=0 \ .
\label{SEforpsiwkb}\end{equation}
Then note that by \eqref{psiWKB}, we have
\begin{equation}
\psi_{WKB}^{-1}\frac{d^2}{dx^2}\psi_{WKB}=\frac{P^2}{\hbar^2}+\frac{1}{2}\{S,x\} \ ,
\end{equation}
so that \eqref{SEforpsiwkb} is equivalent to
\begin{equation}
\frac{P^2}{2m}+\frac{\hbar^2}{4m}\{S,x\}+V-E+X=0 \ .
\label{qurt}\end{equation}
On the other hand, by $P^2=2m(E-V)$, we get
\begin{equation}
X=-\frac{\hbar^2}{4m}\{S,x\} \ ,
\end{equation}
so that
\begin{equation}
\left(-\frac{\hbar^2}{2m}\frac{d^2}{dx^2}+V-E-
\frac{\hbar^2}{4m}\{S,x\}\right)\psi_{WKB}=0 \ .
\label{SchrDaWKB}\end{equation}
Thus, the exact Schr\"odinger equation follows by adding $\hbar^2\{S,x\}/(4m)$ inside the round brackets in \eqref{SchrDaWKB}. Similarly, for $k=\pm 1$, the equation whose solutions in the first-order WKB approximation reproduce \eqref{lesoluzioni} is obtained by adding $-\{\eta_k,t\}/2$ inside the square brackets of \eqref{lineqbhalf}. Thus, by \eqref{421}, the quantum cosmological equations are
\begin{equation}
\left(\frac{d^2}{dt^2}+\frac{\kappa}{4a_k^2(t)}\right) \Psi(t)=0 \ ,
\quad k=\pm 1 \ ,
\label{lineqbhalfquantumdue}\end{equation}
where $a_k(t)$ denotes the explicit expression in terms of
$t$, obtained by solving the Friedmann equations or, equivalently, Eqs.~\eqref{lineqbhalf} and \eqref{F2bis}.
The solutions of \eqref{lineqbhalfquantumdue} define
the quantum scale factor $a_{qk}(t)$ by
\begin{equation}
\Psi_k=\sqrt{a_{qk}}e^{-\frac{i}{2}\sqrt{\kappa}\eta_{qk}} \ , \qquad
\Psi^D_k=\sqrt{a_{q k}}e^{\frac{i}{2} \sqrt{\kappa}\eta_{qk}} \ ,
\label{quelleesatte}\end{equation}
where $\eta_{qk}$ is the quantum conformal time
\begin{equation}
\eta_{qk}(t)=\int_t^{t_0}\frac{dt'}{a_{qk}(t')} \ .
\end{equation}
Interestingly, in the case $k=0$, scale invariance of Friedmann's equations allows $\alpha$ in \eqref{generale} to remain as a free parameter, so that
\begin{equation}
\left(t_P^2\frac{d^2}{dt^2} \pm\frac{\beta^2}{4a_0^2}\right) \Psi = 0 \ ,
\label{quantumFEA}\end{equation}
$\beta\in \RR$, with solutions
\begin{equation}
\Psi_0=\sqrt{a_{q0}}e^{-b\eta_{q0}/t_P} \ , \qquad \Psi_{q0}^D=\sqrt{a_{q0}}e^{b\eta_{q0}/t_P} \ ,
\label{casozero}\end{equation}
with either $b=\beta/2$ or $b=i\beta/2$.
Setting
\begin{equation}
\Psi=e^{\frac{i\sqrt k}{2t_P}\sigma_k} \ ,
\end{equation}
$k=\pm 1$, one sees that the WKB expansion
\begin{equation}
\sigma_k=\sum_{j\geq 0}\left(\frac{t_P}{i\sqrt k}\right)^j\sigma_{kj} \ ,
\end{equation}
 yields, at first-order in $t_P$,
\begin{equation}
\sigma_k=\eta_k-it_P\log \dot\eta_k \ .
\end{equation}
Thus, as is evident
by construction, \eqref{quelleesatte} reduces to \eqref{lesoluzioni} within this
approximation.
Similarly, one can apply this analysis to $\Psi_0$ and $\Psi_0^D$, which then reduce to \eqref{casozero} with $a_{q0}$ and $\eta_{q0}$ replaced by $a_0$ and $\eta_0$, respectively.
Note that we used the WKB expansion of $\sigma$ to verify that, to first order, $a_{qk}$ coincides with $a_k$. However, just as in quantum mechanics, when the energy depends on $\hbar$, the term-by-term comparison in powers of $t_P$ breaks down once we impose the on-shell condition. A related issue
concerns Eqs.~\eqref{QFT} and \eqref{KG}.

\noindent
One may easily verify that the
QSHJEs associated to Eqs.~\eqref{lineqbhalfquantumdue} and \eqref{quantumFEA} read
\begin{equation}
\frac{1}{a_{qk}^2}-\frac{1}{a_k^2}+\frac{2}{\kappa}\{\eta_{qk},t\}=0 \ , \qquad k=\pm 1 \ ,
\label{moltobella1}\end{equation}
and, for $k=0$,
\begin{equation}
\frac{1}{a_{q0}^2}-\frac{1}{a_0^2}\pm\frac{2t_P^2}{\beta^2}\{\eta_{q0},t\}=0 \ .
\label{moltobella2}\end{equation}
Note that, even in this case, $a_k(t)$, $k=-1,0,1$, is intended to be expressed on shell, that is in terms of its explicit time dependent expression,
solutions of the Friedmann's equations or, equivalently, of Eqs.~\eqref{lineqbhalf} and \eqref{F2bis}.

\noindent
The above derivation extends to the case of the momentum $P_k$. Namely, by \eqref{lineqbhalfquantumdue} we have
\begin{equation}
\left(\hbar^2\frac{d^2}{dt^2}+kP_k^2c^2\right) \Phi(t)=0 \ ,
\quad k=\pm 1 \ ,
\label{lineqbhalfquantumdueperP}\end{equation}
where $P_k(t)$ denotes the explicit expression in terms of
$t$, obtained by solving the Friedmann equations. The solutions of
\eqref{lineqbhalfquantumdueperP} read
\begin{equation}
\Phi_k=\frac{1}{\sqrt {P_{qk}}}e^{-i\frac{c\sqrt k}{\hbar}\int_t^{t_0} P_{qk}} \ , \quad \Phi_k^D=\frac{1}{\sqrt {P_{qk}}}
e^{i\frac{c\sqrt k}{\hbar}\int_t^{t_0} P_{qk}} \ ,
\label{lesoluzioniQM+1perP}\end{equation}
$k=\pm 1$.

\section{Phases of quantum curvature}

A key property of Friedmann's equations is that,
for $k\neq0$, the $k$-factor can be absorbed
through the rescaling $a_k\rightarrow a:=a_k/\sqrt \kappa$.
The lack of explicit $k$-dependence of $\rho$ and $p$ implies that $a$ is
also independent of $k$. By $a_{-1}=ia_1$ it follows that
when $a_1$ is real, $a_{-1}$ is purely imaginary, and vice versa. Notably,
Eq.~\eqref{lineqbhalfquantumdue} is equivalent to
\begin{equation}
\left(\frac{d^2}{dt^2}+\frac{1}{4 a^2}\right) \Psi=0 \ ,
\label{lineqbhalfquantumduebisse}\end{equation}
that, by
\begin{equation}
{}^3R=6\kappa/(c a_k)^2=6/(ca)^2 \ ,
\end{equation}
 takes the suggestive form
\eqref{bellissimaFrankZappa}.
For any given solution of \eqref{lineqbhalfquantumduebisse}, the corresponding quantum scale factors are still related by $a_{q-1} = i\,a_{q1}$.
This allows
for continuous transitions between different geometries, as we will show with an example.
In contrast to the classical Friedmann case, real physical solutions exist for both $a_{q1}$ and $a_{q-1}$ for all $t \in \mathbb{R}$, each corresponding to a distinct solution of Eq.~\eqref{lineqbhalfquantumduebisse}.
This mechanism relates to the QSHJE, where the quantum potential ensures the conjugate momentum remains real, even in classically forbidden regions \cite{Faraggi:1997bd,Faraggi:1998pd}.

\section{The quantum radiation era}

\begin{figure}[ht]
\begin{subfigure}[t]{0.5\textwidth}
  \includegraphics[scale=0.65]{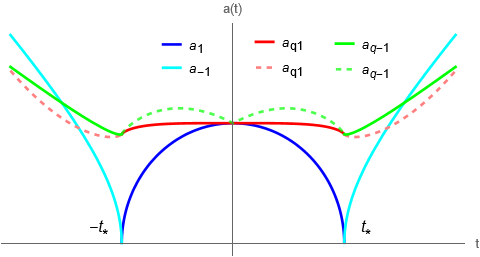}
\end{subfigure}
\begin{subfigure}[t]{0.5\textwidth}
  \includegraphics[scale=0.65]{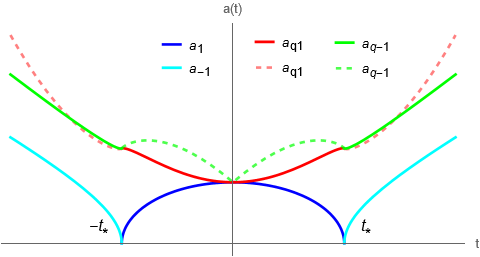}
\end{subfigure}
\begin{subfigure}[t]{0.5\textwidth}
  \includegraphics[scale=0.65]{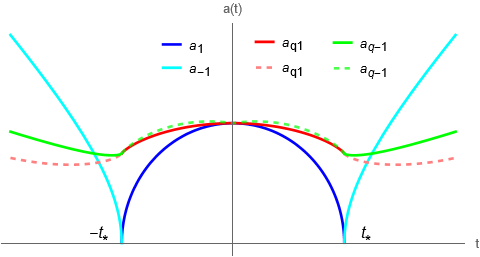}
\end{subfigure}
\caption{$a_k$ versus $a_{qk}$ for the radiation solution.} \label{3plot}
\end{figure}

We now turn to the quantum radiation era (characterized by $p = c^2 \rho/3$ and $\Lambda = 0$), a scenario whose solutions exhibit further intriguing features.
In this case, we have $\rho(t) = \rho(t_0)\,\frac{a_k^4(t_0)}{a_k^4(t)}$, where
\begin{equation}
  a (t):=\frac{a_k(t)}{\sqrt\kappa} = \sqrt{t_*^2-t^2} \ ,
\label{sola1}\end{equation}
and $8\pi G\rho(t_0) a_k^4(t_0)=3 t_*^2/t_P^4$.
 If $k=1$, the solution is valid within the bounded region $|t| < t_*$. For $t<0$, there is an expansion, starting from an initial singularity, $a_1(-t_*)=0$, a ``big bang'', followed by a contraction when $t>0$, leading to a ``big crunch'' at another singular point, $a_1(t_*)=0$. This behavior is represented in Fig. \ref{3plot} by the continuous blue curve. If $k=-1$, the scale factor is given by $a_{-1}=i a_{1}$. Here, the solution applies to the disjoint regions $|t|>t_*$, represented by the two cyan-colored lines.

\noindent
In the case of \eqref{sola1}, Eq.~\eqref{lineqbhalfquantumdue} is equivalent to ($t_*=1$)
\begin{equation}
\left[\frac{d^2}{dt^2}+\frac{1}{4(1-t^2)}\right]\Psi=0 \ .
\label{SW}\end{equation}
The solutions are expressed in terms
of Gauss hypergeometric functions ${}_2F_1$, see \cite{Duverney,Sueishi:2020rug}
for an excellent account on related topics. In the region $t\in [-t_*,t_*]$
\begin{equation} \label{psigenq}
\begin{split}
  \Psi & =  A\, {}_2F_1\left(\frac{\sqrt{2}-1}{4},-\frac{\sqrt{2}+1}{4};\frac{1}{2};\frac{t^2}{t_*^2}\right) \\
  & + B \, t  \; {}_2F_1\left(\frac{\sqrt{2}+1}{4},\frac{1-\sqrt{2}}{4};\frac{3}{2};\frac{t^2}{t_*^2}\right),
\end{split}
\end{equation}
while for $|t|>t_*$,
\begin{equation} \label{psigenq2}
\begin{split}
   \Psi & =  C|t|^{\frac{1}{2} \left(1-\sqrt{2}\right)}\, {}_2F_1\left(\frac{\sqrt{2}-1}{4} ,\frac{\sqrt{2}+1}{4};1+\frac{1}{\sqrt{2}};\frac{t_*^2}{t^2}\right) \\
  & + D|t|^{\frac{1}{2} \left(\sqrt{2}+1\right)} \,  {}_2F_1\left(\frac{-\sqrt{2}-1}{4},\frac{1-\sqrt{2}}{4} ;1-\frac{1}{\sqrt{2}};\frac{t_*^2}{t^2}\right),
\end{split}
\end{equation}
with $A,B,C,D$ complex constants. All ${}_2F_1$'s are real in their domains. In each region we construct the $a_{qk}$, whose square
root appears in the wave function, following the process described in \cite{Faraggi:1997bd} and in sect.14.4 of \cite{Faraggi:1998pd}.

\noindent
We are here interested in deriving the quantum modified scale factor $a_{qk}$ and compare with the $a_k$ of Eq.~\eqref{sola1}, which satisfies the two Friedmann equations \eqref{Freqs} in the case of radiation.
Now that we have expressed the solutions of the quantum equation \eqref{SW} in terms of real, linearly independent functions, we need only use Eqs.~\eqref{a1} and \eqref{aqm1} to reverse-engineer $a_{qk}$. The difference now is that the $\phi_k$ and $\phi_k^D$ here correspond to the linearly independent branches of the solutions \eqref{psigenq} and \eqref{psigenq2}. Thus, we write
\begin{equation} \label{aqex1}
a_{q1} = \frac{|\phi_1^D-i \ell \phi_1|^2}{W t_P  (\ell+\bar{\ell})} \ ,
\end{equation}
for $k=+1$, and
\begin{equation} \label{aqmex1}
a_{q-1} = \frac{\left(\alpha \phi_{-1}^D + \zeta \phi_{-1}\right)\left( \gamma \phi_{-1}^D + \delta \phi_{-1} \right)}{W t_P   \left(\zeta\gamma-\alpha \delta\right)} \ ,
\end{equation}
for $k=-1$, with the $\phi_k$ and $\phi_k^D$ being given in terms of the hypergeometric functions that solve Eq.~\eqref{SW}.

\noindent
In Fig.~\ref{3plot}, we plot $a_{\pm 1}$ and $a_{q\pm 1}$ for various linear combinations of hypergeometric functions. Some of the integration constants are fixed by requiring continuity at $\pm t_*$. The rest of the constants allow the quantum solutions to reveal a richer framework, offering deeper insights into spacetime and dynamics. In particular, the singularity $a_k(\pm t_*) = 0$ is replaced by $a_{qk}(\pm t_*)> 0$.
In many cases of interest, singularities do not arise, as will be shown in an upcoming paper. Moreover, quantum solutions extend into regions where $a_k$ corresponds to Euclidean-signature spacetime (dashed lines).

\section{Cosmological scenario and Seiberg-Witten approach}

Eq.~\eqref{lineqbhalfquantumdue} is defined via solutions of the Friedmann equations. Given the variety of cosmological models, analyzing this equation requires diverse techniques. These often involve topological configurations, drawing parallels to particle theory phenomena like instantons and monopole condensation.
A notable example is the rescaling
\begin{equation}
a\to ia \ ,
\label{complextransform}
\end{equation}
that by  \eqref{lineqbhalfquantumduebisse} and \eqref{SW}
maps $(1-t^2)$ into $(t^2-1)$, yielding a dual differential equation matching the one governing the moduli space of quantum vacua in SW theory \cite{Seiberg:1994rs}
\begin{equation}
\left[\frac{d^2}{du^2}-\frac{1}{4(1-u^2)}\right]a=0 \ ,
\label{SW3}\end{equation}
that follows from the uniformization equation for the thrice-punctured Riemann sphere $\Sigma_{0,3}$ \cite{Matone:1995rx}.
Here, $t$ is replaced by the quantum vacuum parameter $u := \langle \mathrm{Tr}(\phi^2)\rangle$, where $\phi$ is the Higgs field, and we define $\langle \phi \rangle = a\,\sigma^3/2$ with $\sigma^3 = \mathrm{diag}(1,-1)$.
The map \eqref{complextransform} can be seen as either the complex transformation $a_k\to ia_k$, equivalent to a Wick rotation to Euclidean signature, or as curvature interchange $k\to -k$. Thus, in the Friedmann equations, a Wick rotation of the FLRW metric corresponds to scalar curvature interchange. This insight appears linked to the work of Kontsevich, Segal, and Witten \cite{Kontsevich:2021dmb,Witten:2021nzp}.

\noindent The above analysis implies that $t$ and $\Psi$ of the dual quantum Friedmann equation \eqref{SW3} satisfy the relation \cite{Matone:1995rx}
\begin{equation}
t=\pi i \left({\cal F}-\frac{\Psi}{2}\frac{d\cal F}{d\Psi}\right) \ ,
\label{uFrelation}\end{equation}
where the prepotential $\cal F$ is defined by $\Psi^D=\partial_\Psi {\cal F}$.
Here, $\Psi$ and $\Psi^D$ are linearly independent solutions of Eq.~\eqref{SW3}, denoted in SW approach by $a$ and $a_D$, respectively. Furthermore,
$t(\Psi)$ is $\Gamma(2)$ invariant
\begin{equation}
t(\tilde\Psi)=t(\Psi) \ ,
\label{eccellente}\end{equation}
\begin{equation}
\begin{pmatrix} \tilde\Psi^D \\ \tilde\Psi \end{pmatrix}=
\begin{pmatrix} a & b \\ c & d \end{pmatrix} \begin{pmatrix}\Psi^D \\ \Psi \end{pmatrix} \ , \qquad
\begin{pmatrix} a & b \\ c & d \end{pmatrix} \in \Gamma(2) \ ,
\end{equation}
with $a, d \equiv 1 \pmod 2$ and $b, c \equiv 0 \pmod 2$.
$\Gamma(2)$ is the uniformizing group of the thrice-punctured Riemann sphere, i.e., $\Sigma_{0,3}\cong\mathbb{H}/\Gamma(2)$, where $\mathbb{H}$ is the upper half-plane \cite{Matone:1995rx}.
The $\Gamma(2)$ invariance of the dual quantum Friedmann equation \eqref{SW3} suggests that cosmic time evolution may be periodic.
This raises the possibility that observed structures, like the Big Ring \cite{Lopez:2024rzp}, are related to uniformization theory at a fundamental level.
For an introduction to uniformization and Liouville theories, see \cite{Matone:1993tj}, and for discussions in the SW approach, refer to \cite{Matone:1995rx,Matone:1995jr,Matone:1996bj,Bonelli:1996ry,Manschot:2021qqe,Aspman:2021vhs,Aspman:2023ate,Aspman:2022sfj}.
Recently, relation \eqref{uFrelation} and its generalization \cite{Flume:2004rp} have been applied in various contexts, including black hole studies \cite{Aminov:2020yma,Bianchi:2021xpr,Bonelli:2021uvf,Bianchi:2021mft,Fioravanti:2021dce,HuHu,Dodelson:2022yvn,Consoli:2022eey,Fioravanti:2022bqf,%
Bhatta:2022wga,Bianchi:2022qph,Aminov:2023jve,BarraganAmado:2023apy,Bautista:2023sdf,Bianchi:2023sfs,Giusto:2023awo,Ge:2024jdx,Cipriani:2024ygw,%
Arnaudo:2024rhv,Bianchi:2024vmi,Bautista:2024agp,Jia:2024zes,Aminov:2024mul,BarraganAmado:2024tfu,Bautista:2024emt} and resurgence theory \cite{Basar:2017hpr}.

\section{Conclusion and outlook}

\noindent
Unlike the second Friedmann equation, the first one is nonlinear. However, in \cite{Matone:2021rfj}, it was shown that the first Friedmann equation can be reformulated as a linear equation. This result resolves, within this context, a conceptual discrepancy between GR and QM: while the former is highly nonlinear, the latter is intrinsically linear.

\noindent
Here, we have shown that the redshift implies that this equation corresponds to the first-order WKB expansion of the full quantum equation \eqref{lineqbhalfquantumdue} for $k = \pm 1$ and \eqref{quantumFEA} for $k=0$. Remarkably, \eqref{quantumFEA} retains the structure of \eqref{lineqbhalfquantumdue} but with a free parameter, suggesting possible maps between different curvatures.

\noindent
These results highlight that GR and QM emerge from a broader framework that, in the cosmological context, is governed by Eqs.~\eqref{lineqbhalfquantumdue} and \eqref{quantumFEA}, deepening our understanding of fundamental physics. Its link to quantum Hamilton-Jacobi theory, closely tied to GR \cite{Faraggi:1997bd,Faraggi:1998pd,Faraggi:1998va,Faraggi:1998pc,Faraggi:1998pb,Bertoldi:1999zv,Matone:2000ge,Faraggi:2012fv,Faraggi:2020blm}, is particularly significant.
This framework opens new avenues of research. In particular, it implies that multi-fluid cosmologies \cite{mult1,mult2} and the universe’s evolution should be described within the quantum cosmological equations \eqref{lineqbhalfquantumdue} and \eqref{quantumFEA}. This follows from the fact that multiple fluids can be treated as a single effective fluid \cite{mult0}, making it compatible with our approach.

\noindent
Moreover, the modified quantum scale factor may provide insights into unexpected experimental results, such as the observed tension in the universe’s expansion rate.
Note that certain gravitational theories are also used to model quantum effects. In some modifications to GR (e.g. dilaton and Weyl gravity), the conformal invariance is restored at trans-Planckian energies, with the symmetry then spontaneously broken in analogy with the Higgs mechanism
\cite{Slagter:2022srb}.

\noindent
Our findings emphasize the relevance of studying this structure within Einstein’s field equations for general metrics and matter content. It would be interesting to explore its implications in the Wheeler-DeWitt quantization framework \cite{DeWitt}, widely applied in cosmology \cite{wdw1,wdw2,wdw3,wdw4,wdw5,wdw6}, including mini-superspace approximations and the quantization of low-dimensional gravitational theories \cite{wdwl0,wdwl1,wdwl2,wdwl3,wdwl4,wdwl5}.

\noindent
Finally, it should be noted that, in our investigation, the Schwarzian derivative plays a crucial role.
This is closely related to the idea that, like GR, QM also has a geometrical origin.
In particular, as we said, the QSHJE is derived by imposing the existence
of point transformations that connect different states.  This leads to a cocycle
condition that uniquely determines the Schwarzian derivative and then the QSHJE
(see the theorem in Section 9.2 of \cite{Faraggi:1998pd}).
Remarkably, the Schwarzian derivative also appears in related contexts such as the Jackiw–Teitelboim gravity \cite{JT1,JT2,JT3} and the Sachdev–Ye–Kitaev model \cite{SYK1,SYK2,SYK3,SYK4,SYK5,SYK6}.

\noindent

\noindent
We thank Giulio Bonelli, Syo Kamata, and Marco Peloso for valuable discussions. Special thanks to Daniel Duverney for assistance in solving \eqref{SW}.

\end{document}